# On the Algorithmic Complexity of the Mastermind Game with Black-Peg Results

Michael T. Goodrich

Dept. of Computer Science and

Secure Computing and Networking Center

University of California, Irvine

http://www.ics.uci.edu/~goodrich/

**Abstract**

In this paper, we study the algorithmic complexity of the Mastermind game, where results are single-color black pegs. This differs from the usual dual-color version of the game, but better corresponds to applications in genetics. We show that it is NP-complete to determine if a sequence of single-color Mastermind results have a satisfying vector. We also show how to devise efficient algorithms for discovering a hidden vector through single-color queries. Indeed, our algorithm improves a previous method of Chvátal by almost a factor of 2.

## 1 Introduction

Mastermind [2, 4] is a game played between two players—a *codemaker* and a *codebreaker*—using colored pegs. (See Figure 1.)

Viewed mathematically, Mastermind is abstracted as a game where the codemaker selects a plaintext vector, $V$, of length $N$, whose elements are selected from an alphabet of size $K$. For consistency with the board game, the members of this alphabet are often referred to as "colors." The codemaker and codebreaker both know the values of $N$ and $K$, and game play consists of the codebreaker repeatedly making guesses, $V_1, V_2, \ldots$, about the identity of $V$. For each guess $V_i$, the codemaker provides a score on how well $V_i$ matches $V$. In *double-count* Mastermind, which is the standard version based on the board game, this score consists of a pair of two numbers:

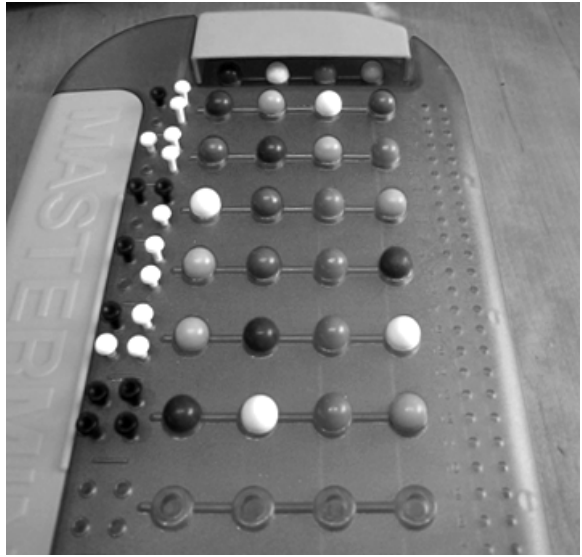

Figure 1: The Mastermind game. The four large pegs in the middle are used for guessing. The four smaller peg locations on the left are used to score each guess—with black-peg and white-peg scores. (Image, Copyright 2009, Michael T. Goodrich. Used with permission.)

- A *black* count, $b(V_i)$, which is the number of elements in $V_i$ and $V$ that match in both value and location. That is,
$$b(V_i) = |\{j\colon V_i[j] = V[j]\}|.$$

- A *white* count, $w(V_i)$, which is the number of elements in $V_i$ that appear in $V$ but in different locations than their locations in $V_i$. That is, letting $\pi$ denote an arbitrary permutation,
$$w(V_i) = \max_\pi |\{j\colon V_i[\pi(j)] = V[j]\}| \; - \; b(V_i).$$

In *single-count* Mastermind, which has been less studied, the codebreaker is given only the black-peg count, $b(V_i)$, for each guess, $V_i$. (Note that it is impossible to solve the problem given only white-count scores.) The goal is for the codebreaker to discover $V$ using a small a number of guesses.

## 1.1 Previous Related Work

The original Mastermind game was invented in 1970 by Meirowitz, as a board game having holes for sequences of length $N = 4$ and $K = 6$ colored pegs. Knuth [4] subsequently showed that this instance of the Mastermind game can be solved in five guesses or less. Chvátal [2] studied the combinatorics of general Mastermind, showing that it can be solved in polynomial time, in the $K \geq N$ case, using $2N\lceil \log K \rceil + 4N$ guesses, and Chen *et al.* [1] showed how this bound can be improved, in this case, to $2N\lceil \log N \rceil + 2N + \lceil K/N \rceil + 2$ guesses. Stuckman and Zhang [6] showed that it is NP-complete to determine if a sequence of guesses and responses in general double-count Mastermind is satisfiable.

## 1.2 Our Results

In this paper we study the single-count (black-peg) version of *Mastermind*. Such a scenario is motivated from genomic data, where a genomic database owner, Dave, can "play" a type of Mastermind game with a genomic query string $Q$–for which a querier thinks that he is querying Dave in a privacy-preserving manner—but instead Dave is discovering the full identity of $Q$. That is, $Q$ is iteratively compared with strings provided by Dave (assumed to be from his database, $D$), with each done in a privacy-preserving online manner, so that all is learned from each comparison is the score measuring the similarity of the two strings, with the (black-peg) score for each string comparison being revealed to the database owner (and possibly also the owner of $Q$) before the next comparison begins.

We begin our discussion by showing that, in fact, the problem of determining whether a sequence of Mastermind responses has a valid solution is NP-complete, even if each response is a single-count response. In addition to the NP-completeness result, we show that an arbitrary query string, $Q$, of length $N$ from an alphabet of size $K$, can be discovered with $N\lceil \log K \rceil + \lceil (2 - 1/K)N \rceil + K$ guesses, each of which is a single-count response. This improves the Chvátal upper bound by almost a factor of $2$.

## 2 Black-Peg Mastermind is NP-Complete

As mentioned above, Stuckman and Zhang [6] show that double-count Mastermind satisfiability is NP-complete. Unfortunately, their proof, which is based on a reduction from the well-known Vertex Cover problem, does not translate into a proof that single-count Mastermind satisfiability is NP-complete. So we provide such a proof in this section. The implications of this fact are that satisfying an arbitrary sequence of Mastermind queries should be considered computationally infeasible.

In the single-count Mastermind satisfiability problem, we are given a sequence of Mastermind queries, $V_1, V_2, \ldots, V_N$, and the responses, $b(V_1), b(V_2), \ldots, b(V_N)$, each of which is said to report the number of indices such that the characters in a $V_i$ and an unknown vector, $V$, at these locations match. We are asked to determine if there indeed exists a vector $V$ that satisfies all of these responses.

**Theorem 1:** *Single-count Mastermind satisfiability is NP-complete.*

**Proof:** It is easy to see that single-count Mastermind satisfiability is in NP. For example, we could nondeterministically guess a vector $V$ and then test in polynomial time whether it satisfies all the responses, $b(V_1), b(V_2), \ldots, b(V_N)$.

To prove that single-count Mastermind satisfiability is NP-hard, we provide a reduction from 3-Dimensional Matching (3DM), which is a well-known NP-complete problem (e.g., see [3]). In the 3DM problem, we are given three sets,

$$X = \{x_1, \ldots, x_n\}, \ Y = \{y_1, \ldots, y_n\}, \ \text{and} \ Z = \{z_1, \ldots, z_n\},$$

of $n$ elements each. In addition, we are given a set $T$ of $m$ triples, $\{(x_{i_1}, y_{j_1}, z_{k_1}), \ldots, (x_{i_m}, y_{j_m}, z_{k_m})\}$, whose elements are respectively taken from the three sets, $X$, $Y$, and $Z$. The problem is to determine if there is a subset of triples such that each element in $X$, $Y$, and $Z$ appears in exactly one triple in this subset.

Suppose, then, that we are given an instance of the 3DM problem, as described above. We consider the unknown vector, $V$, to consist of the following sequence of variables:

$$(X_1, \ldots, X_n; Y_1, \ldots, Y_n; Z_1, \ldots, Z_n; T_1, \ldots, T_m),$$

where the semi-colons are used for the sake of notation to separate the four sections in the unknown vector, $V$. We perform our reduction by constructing a sequence of guess vectors, $V_1, V_2, \ldots, V_N$, together with their responses, $b(V_1), b(V_2), \ldots, b(V_N)$, so that there is a satisfying vector $V$ for these responses if and only if there is a solution to the given instance of the 3DM problem. Our construction begins by setting the number of colors, $K$, to be $m + 1$. Intuitively, there is a color associated with each triple in $T$, plus a "null" color, $\phi$, which is guaranteed not to appear in our unknown vector, $V$. We begin our sequence of queries with three special "enforcer" queries:

$$V_1 = (\phi, \ldots, \phi; \phi, \ldots, \phi; \phi, \ldots, \phi; \phi, \ldots, \phi),$$

which has response $b(V_1) = 0$,

$$V_2 = (\phi, \ldots, \phi; \phi, \ldots, \phi; \phi, \ldots, \phi; 1, 1, \ldots, 1),$$

which has response $b(V_2) = n$, and

$$V_3 = (\phi, \ldots, \phi; \phi, \ldots, \phi; \phi, \ldots, \phi; 0, 0, \ldots, 0),$$

which has response $b(V_3) = m - n$. Intuitively, $V_1$ enforces the fact that the null color, $\phi$, cannot appear in the unknown vector, $V_2$ enforces a counting rule that exactly $n$ of the $T_i$'s will be set to 1, and $V_3$ enforces a counting rule that the remaining $m - n$ of the $T_i$'s will be set to 0. For each triple, $T_s = (x_{i_s}, y_{j_s}, z_{k_s})$, we construct three query vectors, as follows.

$$V_{s,1} = (\phi, \ldots, \phi, s, \phi, \ldots, \phi; \phi, \ldots, \phi; \phi, \ldots, \phi; \phi, \ldots, \phi, 0, \phi, \ldots, \phi),$$

where the $s$ is in position $i_s$ in the first group and the 0 is in position $s$ in the fourth group. This vector has response, $b(V_{s,1}) = 1$. Next, we construct

$$V_{s,2} = (\phi, \ldots, \phi; \phi, \ldots, \phi, s, \phi, \ldots, \phi; \phi, \ldots, \phi; \phi, \ldots, \phi, 0, \phi, \ldots, \phi),$$

where the $s$ is in position $j_s$ in the second group and the 0 is in position $s$ in the fourth group. This vector has response, $b(V_{s,2}) = 1$. Finally, we construct

$$V_{s,3} = (\phi, \ldots, \phi; \phi, \ldots, \phi; \phi, \ldots, \phi, s, \phi, \ldots, \phi; \phi, \ldots, \phi, 0, \phi, \ldots, \phi),$$

where the $s$ is in position $k_s$ in the third group and the 0 is in position $s$ in the fourth group. This vector has response, $b(V_{s,3}) = 1$. Intuitively, these three responses collectively form a "chooser" gadget, where we will either have $T_s = 0$ or the three variables $X_{i_s}$, $Y_{j_s}$, and $Z_{k_s}$, will each be set to have color $s$ (and $T_s = 1$).

This reduction can clearly be done in polynomial time. So all that remains is for us to show that it works. Suppose, then, that there is a possible solution to the given instance of 3DM. Then for each chosen triple, $T_s = (x_{i_s}, y_{j_s}, z_{k_s})$, we can assign colors $T_s = 1$, $X_{i_s} = s$, $Y_{j_s} = s$, and $Z_{k_s} = s$, which will satisfy each of the $V_{s,1}$, $V_{s,2}$, and $V_{s,3}$ vector responses for this value of $s$. Likewise, setting $T_s = 0$ will satisfy each of the $V_{s,1}$, $V_{s,2}$, and $V_{s,3}$ vector responses for a triple $T_s$ that is not chosen. Finally, given that there are $n$ chosen vectors, we will satisfy the three preliminary vector responses as well.

Suppose, alternatively, that we have a vector $V$ that satisfies all of our vector responses. We know that each $X_i$, $Y_j$, and $Z_k$ must be assigned a color other than $\phi$. Since there are only $m + 1$ colors, this implies each $X_i$, $Y_j$, and $Z_k$ must be assigned a color corresponding to a triple number, $s$. If this $T_s = 1$, then in order to have satisfied the vectors $V_{s,1}$, $V_{s,2}$, and $V_{s,3}$, we must have set $X_{i_s} = s$, $Y_{j_s} = s$, and $Z_{k_s} = s$, which implies we can include the triple $(X_{i_s}, Y_{j_s} Z_{k_s})$ in our matching. If $T_s = 0$, then we do not include this triple in our matching. By the vector responses $V_2$ and $V_3$, we know that the number of triples chosen in this way is exactly $n$. Thus, we have found a valid 3-dimensional matching. ∎

Thus, it is extremely unlikely that we will be able to find a polynomial-time algorithm that can solve arbitrary Mastermind query sequences, even if they are single-count results. Note that this is not the same as a guarantee that discovering a string $Q$ requires a long query sequence, however. For we show, in the section that follows, that such query strings, $Q$, can be discovered fairly efficiently using a single-count Mastermind algorithm.

## 3 A Mastermind Algorithm for Single-Count Match Queries

In this section, we explore an algorithm for the single-count Mastermind game, where the code-breaker, Dave, engages in a series of guesses against the unknown string, $Q$, each of which reveals only the single-count score between the query string $Q$ and strings provided by Dave, in an iterative online fashion. Here, we show that Dave can learn $Q$ with a sequence of $N \lceil \log K \rceil + \lceil (2 - 1/K)N \rceil + K$ guesses, where $N$ is the length of $Q$ and $K$ is the size of the alphabet (whose members we call "colors").

We begin the algorithm for Dave by having him perform $K$ queries, each of which is a vector of elements that are all the same color. This allows us to initially know the cardinality, $c_1, c_2, \ldots, c_K$, of every color in the unknown vector, $Q$. If any $c_i = 0$, then we remove the color $i$ from our alphabet of colors, and update the value of $K$ accordingly. The remainder of Dave's computation proceeds as a recursive divide-and-conquer algorithm, which is similar in structure to the approach of Chvátal [2], but improves his bound by almost a factor of 2, even though his algorithm was for the general two-color case, by reusing knowledge gained in previous reclusive calls.

The generic problem is to determine the values of all the elements in a range $Q[l..r]$, which initially is the entire vector $Q = Q[0..N-1]$, assuming we know the values of $c_1, c_2, \ldots, c_K$, of every color in $Q[l..r]$, and each $c_i > 0$. If $K \leq 1$, we are done; so let us assume without loss of generality that $K \geq 2$. In addition, we assume inductively that we know $d$, the number of instances of color 1 outside of the range $Q[l..r]$. Initially, of course, $d = 0$.

Given this initial setup, we split $Q[l..r]$ into $Q[l..m]$ and $Q[m+1..r]$, where $m$ is in the middle of the interval $[l, r]$. The main challenge, then, is to provide for $Q[l..m]$ and $Q[m+1..r]$ the same setup we had for $Q[l..r]$. This setup can be accomplished by determining the cardinalities, $x_1, x_2, \ldots, x_K$ and $y_1, y_2, \ldots, y_K$, of every color that respectively appears in $Q[l..m]$ and $Q[m+1..r]$. We do this with a series of $K - 1$ additional queries, where we guess that the elements in $Q[l..m]$ are of color $i$, for $i = 2, 3, \ldots, K$, and that the rest of $Q$ is of color 1. Let the values of these queries be denoted as $b_2, b_3, \ldots, b_K$, and note that, at this point, we know the following:

$$x_i + y_i = c_i, \quad \text{for } i = 1, 2, \ldots, K \tag{1}$$

$$x_i + y_1 = b_i - d, \quad \text{for } i = 2, 3, \ldots, K \tag{2}$$

$$x_1 + x_2 + \cdots + x_K = m - l + 1. \tag{3}$$

Thus, we can determine $y_1$, as

$$y_1 = \frac{c_1 + \sum_{i=2}^{K}(b_i - d) - (m - l + 1)}{k},$$

for $y_1$ is counted $K$ times in the sum of $c_1$ and all the $(b_i - d)$'s, and the sum of the $x_i$'s is $m-l+1$, by Equation (3). Given the value of $y_1$, we can then determine all the $x_i$ values, by using Equation (1) for $x_1$ and Equation (2) for $x_2, x_3, \ldots, x_K$. Moreover, once we have all these $x_i$ values, we can determine the values, $y_2, y_3, \ldots, y_K$, using Equation (1). Finally, we can determine the values $d' = d + y_1$ and $d'' = d_{x_1}$ and use these respectively for the role of $d$ in $Q[l..m]$ and $Q[m+1..r]$. This gives us all the values necessary to then recursively determine $Q[l..m]$ and $Q[m+1..r]$.

Let us, therefore, analyze the number, $G(N, K)$, of vector guesses performed by this algorithm. Ignoring for the time being the initial set of $K$ guesses, we can bound this parameter using the following recurrence:

$$G(N, K) = 2G(N/2, K) + \min\{N, K - 1\}.$$

Thus, adding the initial $K$ queries back in, we get that the total number of guesses is at most

$$N \lceil \log K \rceil + \lceil (2 - 1/K)N \rceil + K.$$

Therefore, we have the following.

**Theorem 2:** *Given an unknown length-$N$ string $Q$, defined on an alphabet of size $K$, a Mastermind algorithm can discover $Q$ in polynomial time using $N \lceil \log K \rceil + \lceil (2 - 1/K)N \rceil + K$ tests against $Q$, each of which reveals only the number of positions where $Q$ and the test string match.*

## 4  Conclusion

We have shown that, even though the single-count and sequence-alignment Mastermind satisfiability problems are NP-complete, one can effectively construct single-count Mastermind algorithms on arbitrary character strings just by knowing basic information about the length of the strings and the number of characters in the alphabet used to construct those strings.

## Acknowledgments

We would like to thank Pierre Baldi, David Eppstein, Daniel Hirschberg, Stas Jarecki, and Michael Nelson for helpful discussions regarding the topics of this paper. This research was supported in part by the National Science Foundation under grants 0724806, 0713046, and 0847968.